# Space-Like Particle Production: an Interpretation Based on the Majorana Equation

*Luca Nanni

*corresponding author e-mail: luca.nanni@student.unife.com

### Abstract

This study reconsiders the decay of an ordinary particle in bradyons, tachyons and luxons in the field of the relativistic quantum mechanics. Lemke already investigated this from the perspective of covariant kinematics. Since the decay involves both space-like and time-like particles, the study uses the Majorana equation for particles with an arbitrary spin. The equation describes the tachyonic and bradyonic realms of massive particles, and approaches the problem of how space-like particles might develop. This method confirms the kinematic constraints that Lemke's theory provided and proves that some possible decays are more favourable than others are.

## 1  Introduction

The study of faster-than-light particles is a branch of theoretical physics still much debated. It leads to speculations and discussions ranging from a purely scientific scope to a metaphysical-philosophical one [1-5]. In the second half of the last century, several physicists developed an intensive effort to extend the theory of relativity. Their goal was to apply it to massive particles travelling at velocities higher than the speed of light. Among these physicists, the names of Recami, Surdashan and Feinberg stand out [5-7]. They introduced the reinterpretation principle, similar to the one Feynman-Stueckelberg proposed to explain the negative energy of antiparticles in quantum field theory. This solved the superluminal propagation dilemma by restoring the principle of causality. Yet things do not go as well when we attempt to introduce the tachyon into quantum field theory. Problems such as vacuum instability and the violation of change, parity and time reversal (CPT) symmetry are hard to solve [6,8-10]. Physicists have solved these issues separately; but so far, there is not a field theory able to explain the quantum behaviour of subluminal and superluminal massive particles without encountering the problems mentioned. For instance, if a theory imposed compliance with the CPT theorem, then the relationship between spin and statistics would be reversed [9]. Instead, the introduction of a privileged reference frame would solve the problem of vacuum instability. However, this would weaken the theory of relativity's second postulate on the equivalence of inertial reference frames [10].

Although theoretical physics is making progress in this field [11], no experiment has ever proved the existence of tachyons directly or indirectly. Considering the high technology of current measuring instruments, the lack of experimental proof strengthens the position of sceptics on the nonexistence of superluminal particles. However, we cannot exclude *a priori* the possibility that phenomena leading to the production of tachyons have a low probability of occurrence and/or take place only in extreme conditions not yet accessible to current measuring apparatuses [12]. These are good reasons to continue the research on tachyon physics. Their experimental confirmation could radically change current cosmological theories such as the inflationary one.

The purpose of this work is to study the decay of an ordinary particle in bradyons, luxons and tachyons within the framework of quantum mechanics. Such a source of superluminal particles could drive experimental research in the right direction. Lemke has investigated the covariant kinematics of this phenomenon [13]. However, a theory for the possible mechanism of this occurrence still does not exist. This work attempts to overcome this lack by using the Majorana equation for particles with an arbitrary spin [14]. It is possible to propose a mechanism of production through the concept of excited state [15]. This avoids the difficulties arising when applying quantum field theory to superluminal particles.

## 2  Methodological Approach

The Majorana equation is a powerful tool for investigating particles with an arbitrary spin. It can help explore phenomena leading to the production of bradyons and tachyons, since it describes the behaviour of both subluminal and superluminal massive particles [14]. For the bradyonic realm, this equation leads to a discrete mass spectrum that depends on the particle's intrinsic angular momentum. When the reference frame is that of the centre of mass, the particle is in the fundamental state. All other states with increasing intrinsic angular momentum have a decreasing mass. Their occupation probability increases with the particle's velocity [15]. The transition from a given quantum state to another with a higher intrinsic angular momentum decreases the particle's rest mass. The transition also emits energy for the production of luxons and tachyons. This is the mechanism proposed for the decay of an ordinary particle that Lemke discussed [13].

Majorana formulated an equation for elementary particles; however, there are no restrictions to apply it to ordinary particles. The scientific literature includes uses of the Majorana equation to investigate composite systems like the hydrogen atom [16]. This also justifies the equation's use in this work. The discussion below reviews in a concise but comprehensive way Lemke's kinematics theory and the discrete mass spectrum obtained by solving the Majorana equation.

## 2.1 Relativistic Kinematics for the Production of Space-Like Particles

Lemke investigated the decay of an ordinary particle with rest mass $M$ in a number $b$ of bradyons with mass $m_k$ ($k = 1, 2, \ldots, b$). He also examined a number $t$ of tachyons with mass $m_l$ ($l = 1, 2, \ldots, t$), and massless luxons [13]. To comply with covariant kinematics, the decay must hold the following constraints:

1) To denote by $\boldsymbol{P}$ the four-momentum of the ordinary particle, and by $\boldsymbol{p}_k$ that of the $k$th tachyon produced, it must satisfy the following constraints:

$$\begin{cases} Pp_l \geq 0 & l = 1, \ldots, t \\ p_k p_l \leq 0 & l, k = 1, \ldots, t \end{cases} \tag{1}$$

In the reference frame of the decaying particle's centre of mass, these constraints ensure that the energies of the tachyons are positive definite. They also ensure that the kinematics of the obtained particles is finite—that there is no singularity concerning momentum and energy.

2) The total momentum of the tachyons produced in the decay must be space-like:

$$\sum_{l=1}^{t} p_l^2 \leq 0 \quad l = 1, \ldots, t \tag{2}$$

The negative value of (2) results from the fact that the tachyonic mass is imaginary. Since the momentum of the decaying bradyon is real and positive, it follows from constraint (2) that a decay that produces only tachyons is not possible.

3) In accordance with constraint (1), the total energy of the produced tachyons is lower bound. It follows that all other particles obtained from the decay have a bounded momentum. Therefore, if $M$ is the rest mass of the decaying particle and $m_k$ ($k = 1, \ldots, b$) is the rest mass of the produced particles, then the maximum possible number of bradyons obtained is:

$$b_{max} = \frac{M}{\sum_{k=1}^{b} m_k} \tag{3}$$

4) The number of tachyons obtained from the decay is limited. If $m_{min}$ is the smallest mass of the produced tachyons, then their maximum number is:

$$t_{max} < \frac{M^2}{|m_{min}^2|} \tag{4}$$

Lemke concluded that to ensure lepton and baryon number conservation, the following constraint must hold:

$$\begin{cases} q_k^2 \geq M^2 \\ -p_l^2 \geq M^2 \end{cases} \tag{5}$$

In this constraint, $q_k$ is the four-momentum of the $kth$ bradyon produced in the decay. In the case of constraint **(5)** above, not all obtained tachyons can have positive definite energy. The latter case can only allow emission of tachyons with no positive energy.

In the next section, we prove that the kinematics constraints of Lemke's theory meet the results obtained by solving Majorana's equation. This demonstrates that the equation is a valid tool to confront the problem of tachyon production.

## 2.2 Bradyons and Tachyons from the Perspective of Majorana's Equation

A relativistic quantum theory that includes tachyons requires an infinite-dimensional representation of the Lorentz group $SO(1,3)$ [7-17]. Majorana formulated his equation using just this algebraical framework [14]. Therefore, it is suitable to study processes where tachyons and bradyons are involved. Majorana's equation is:

$$\left(\mathbb{1}i\hbar\frac{\partial}{\partial t} - \boldsymbol{\alpha}_1 i\hbar\frac{\partial}{\partial x} - \boldsymbol{\alpha}_2 i\hbar\frac{\partial}{\partial y} - \boldsymbol{\alpha}_3 i\hbar\frac{\partial}{\partial z} - \boldsymbol{\beta} m_0 c^2\right)|\Psi\rangle = 0 \quad (6)$$

In this equation, $\boldsymbol{\alpha}_i$ and $\boldsymbol{\beta}$ are infinite matrices. To avoid solutions with negative energy, Majorana's equation requires that $\boldsymbol{\beta}$ must be positive definite. This constraint leads to subluminal solutions with a discrete mass spectrum:

$$m(J_n) = \frac{m_0}{\left(\frac{1}{2}+J_n\right)} \quad (7)$$

In equation **(7)**, $m_0$ is the rest mass of the particle and $J_n$ is the intrinsic angular momentum given by:

$$J_n = s + n \quad n = 1,2,... \quad (8)$$

In **(8)**, $s$ is the spin of the particle (i.e. the particle's intrinsic angular momentum in its fundamental state) and $n$ is the order of the excited state. Equation **(7)** holds both for bosons ($J_n$ is an integer number) and fermions ($J_n$ is a half-integer number). All the excited states have an intrinsic angular momentum $J_n > s$ and have an occupation probability proportional to $(v/c)^n$ [15]:

$$P_n = \sqrt{(v/c)^n - (v/c)^{n+1}} \quad (9)$$

In **(9)**, $v$ is the particle velocity and $c$ is the speed of light. This equation shows that the occupation probability of an excited state increases with the particle's velocity. At a constant speed, the probability decreases with order $n$ of the excited state.

The tachyonic solutions of the Majorana equation instead have a continuous spectrum of mass and energy that can be both positive and negative. In addition, these states may be evident as excited states of the initial particle. In fact, when the particle's velocity approaches the speed of light, the bradyonic mass spectrum tends to become continuous to join up with the tachyonic one.

Because of its space-like solutions, the Majorana equation is not local. Moreover, since the matrix $\beta$ must be positive definite, the CPT theorem does not hold. For instance, one of the transformations this theorem provides is:

$$\gamma^0 \to -\gamma^0 \tag{10}$$

In **(10)**, $\gamma^0 = \beta^{-1} = \beta^\dagger$. However, since $\beta$ must be positive definite, it cannot have negative eigenvalues. Thus, transformation **(10)** is not possible. Using the Majorana equation to investigate the decay of an ordinary particle in bradyons, luxons and tachyons means avoiding the problems arising from quantum field theory extended to superluminal particles.

The solutions of the Majorana equation are compatible with the constraints that Lemke's theory provided. For instance, the bounded momentum of the produced bradyons and their limited number are coherent with their discrete mass spectrum. In fact, given the velocity of the decaying particle, with **(9)** it is possible to evaluate which occupied state is more probable and thus, which is the more probable mass of the produced bradyons. According to this picture, the produced bradyons are not more than excited states of the initial particle with lower mass. The energy emitted in this transition leads to the production of luxons and tachyons.

## 3   Decay Mechanisms of an Ordinary Particle

Let us consider a fermion with rest mass $m_0$ and spin $s$, travelling at a subluminal velocity $u$. We may rewrite the mass spectrum expressed in **(7)** as:

$$m(n) = \frac{m_0}{(n+1)} \quad n = 1,2,\ldots \tag{11}$$

In **(11)**, $n$ is the order of the excited state. The transition from one excited state to the next occurs with a decrease in mass and an increase in the intrinsic angular momentum $J_n$:

$$\Delta m(n \to n+1) = \frac{m_0}{(n+1)(n+2)} \tag{12}$$

The energy produced in this transition is:

$$\Delta E = \frac{\gamma m_0 c^2}{(n+1)(n+2)} \tag{13}$$

Equation **(13)** holds if the produced bradyons have the same velocity of the original particle. In other words, the Lorentz factor does not change after the transition. Let us suppose this energy leads to the production of a tachyon. This must be a space-like solution of the Majorana equation for the original particle:

$$E^2 = p^2 c^2 - m_0^2 c^4 \tag{14}$$

with the constraint $p^2 c^2 > m_0^2 c^4$. Obtaining from **(13)** the term $m_0^2 c^4$ and knowing that $p^2 c^2 = \gamma^2 m_0^2 v^2 c^2$ (where $v$ is the tachyon's velocity), the inequality $p^2 c^2 > m_0^2 c^4$ gives us:

$$v^2 > \frac{\Delta E^2}{\gamma^4 m_0^2 c^2}[(n+1)(n+2)]^2 \tag{15}$$

Making explicit the Lorentz factor and considering that $v > c$, **(15)** becomes:

$$\frac{\Delta E^2 (c^2-u^2)^2}{m_0^2 c^8}[(n+1)(n+2)]^2 > 1 \tag{16}$$

Equation **(16)** yields the maximum value of the initial particle's mass $m_0$ needed because the following decay process takes place:

$$m_0 < \Delta E \frac{(c^2-u^2)}{c^4}(n+1)(n+2) \tag{17}$$

This mass gets smaller and smaller as the velocity $u$ approaches the speed of light. In addition, according to **(9)**, this increases the occupation probability of states with high $n$, with a consequent reduction of amplitude $\Delta E$. If the decay leads also to the production of luxons, then the tachyon's velocity will be lower than **(15)**. Particularly, when $v = c$, the decay produces only bradyons and luxons.

Notably, this mechanism always leads to the production of bradyons with intrinsic angular momentum $J_n$ higher than that of the decaying particle. It never leads to the production of only tachyons. This result is in agreement with the second constraint of Lemke's theory.

The rest mass of the produced bradyon is $m(n+1) = m_0/(n+2)$. Therefore, according to **(3)**, the maximum number of bradyons obtained from the decay is:

$$b_{max} = \frac{m_0}{m(n+1)} = (n+2) \tag{18}$$

The higher the order of the excited state, the greater is the number of bradyons produced.

When $n = 0$, the decaying particle is in the fundamental state and the produced bradyon is the first excited state. With this value, **(13)** yields:

$$\Delta E_0 = \gamma \frac{m_0 c^2}{2} \tag{19}$$

while for other excited states $\Delta E \leq \Delta E_0$. Since this is the energy for the production of tachyons (and eventually luxons), it follows that it is low bounded. This result is in agreement with the first constraint of Lemke's theory.

From **(13)**, it is possible to obtain the mass of the produced tachyon. For this purpose, the energy of a tachyon is [18]:

$$E_t = \frac{m}{\sqrt{\frac{v^2}{c^2}-1}} c^2 sgn\left\{1 - \frac{v \cdot u}{c^2}\right\} \tag{20}$$

The function $sgn$ is the anomalous factor that Park's theory introduced [18] about the relativistic dynamic of tachyons. Therefore, we can write:

$$\Delta E = E_t + h\nu \Rightarrow \frac{m_0 c^2}{\sqrt{\frac{v^2}{c^2}-1}} \frac{1}{(n+1)(n+2)} = \frac{m_t}{\sqrt{\frac{v^2}{c^2}-1}} c^2 sgn\left\{1 - \frac{v \cdot u}{c^2}\right\} + h\nu \tag{21}$$

From this, we get the tachyonic mass:

$$\mu = m_t sgn\left\{1 - \frac{v \cdot u}{c^2}\right\} = \left[\frac{m_0 c^2}{\sqrt{\frac{v^2}{c^2}-1}} \frac{1}{(n+1)(n+2)} - h\nu\right] \frac{\sqrt{\frac{v^2}{c^2}-1}}{c^2} \quad (22)$$

Equation **(22)** proves that not only is the tachyon's energy bounded, but so is that of the possible luxons—there are no singularities concerning the energy. This strengthens the first constraint of Lemke's theory.

Let us consider the case in which the initial particle decays in a bradyon without the production of tachyons and luxons. We are therefore considering a transition of the decaying particle from an excited state $n$ to an excited state $n + k$ ($k = 1, 2, \ldots$). In the hypothesis that the energy is conserved, we can write:

$$\gamma_1 \frac{m_0 c^2}{(n+1)} = \gamma_2 \frac{m_0 c^2}{(n+k+1)} \quad (23)$$

If $u_1$ and $u_2$ are respectively the velocity of the decaying particle and that of the produced bradyon, equation **(23)** gives us:

$$u_2^2 = c^2 - \frac{(n+1)^2}{(n+1+k)^2}(c^2 - u_1^2) \quad (24)$$

Therefore, the energy produced in the decay gives only a kinetic contribution to the obtained bradyon. Its velocity is much closer to the speed of light, as its intrinsic angular momentum $J_{n+k}$ is higher. Notably, if $u_1 \cong c$, then also $u_2$ tends to $c$. In addition, according to **(9)**, the particle tends not to decay. The occupation probability of the produced bradyon compares with that of the initial particle. The eventual production of luxons does not change the mechanism of this possible decay.

Let us generalize the decay mechanism that leads to the production of bradyons and tachyons. In particular, let us suppose that the produced bradyon occupies an excited state with an intrinsic angular momentum $J_{n+k}$ with $k = 2, 3, \ldots$, and that its velocity is different from that of the decaying particle. The energy emitted in the transition from the initial particle to the produced bradyon is:

$$\Delta E = \gamma_2 \frac{m_0 c^2}{(n+k+1)} - \gamma_2 \frac{m_0 c^2}{(n+1)} \quad (25)$$

The obtained energy goes toward the production of the tachyon. Supposing this energy is positive, we can write:

$$m_0 c^2 \left(\frac{\gamma_2}{(n+k+1)} - \frac{\gamma_1}{(n+1)}\right) = \gamma_t m_0 c^2 > 0 \quad (26)$$

In turn, **(26)** gives us:

$$\frac{\gamma_2}{(n+k+1)} - \frac{\gamma_1}{(n+1)} > 0 \quad \Rightarrow \quad \gamma_2 > \gamma_1 \frac{(n+k+1)}{(n+1)} \quad (27)$$

Equation **(27)** holds if velocity $u_2$ of the produced bradyon is lower than velocity $u_1$ of the decaying particle. However, according to **(9)**, the occupation probability of the excited state is lower than that of the decaying particle. Therefore, this mechanism of decay is less favourable than the one leading to the production of a bradyon with a velocity higher than that of the initial particle. Things change if the decay leads to the production of a tachyon with negative energy. In this case, **(27)** becomes:

$$\gamma_2 < \gamma_1 \frac{(n+k+1)}{(n+1)} \qquad (28)$$

that holds if $u_2 > u_1$. In this way, the produced bradyon has an occupation probability greater than that of the decaying particle. Therefore, this mechanism is more favourable than the ones investigated above. The result obtained fulfils the fourth constraint of Lemke's theory.

The investigation performed in this section is also valid in the case of a bosonic particle and must be performed using **(7)** with integer number $J_n$.

## 4  Conclusion

This analysis reviewed the decay of an ordinary particle in bradyons, luxons and tachyons from the perspective of quantum mechanics. It used the Majorana equation for a particle with an arbitrary spin. The mechanism for the decay proposes that:

a) The initial particle must be in a high Majorana excited state, which corresponds to high velocity and high intrinsic angular momentum.

b) The particle decays in a bradyonic excited state with lower mass and higher intrinsic angular momentum. The energy emitted in this transition leads the production of tachyons and/or luxons.

The investigation performed leads to the following results:

1. The proposed mechanism leads always to the production of bradyons. It is never possible to obtain only tachyons, in agreement with Lemke's relativistic kinematics.
2. The energy of the bradyons, luxons and tachyons is always lower bound, avoiding any singularities.
3. The favoured mechanisms are those leading to the production of only bradyons or bradyons and tachyons with negative energy.
4. The production of luxons is possible for any decay mechanism in respect of the principle of energy conservation.

These results confirm Lemke's on the covariant kinematics of an ordinary particle's decay. The approach proposed in this study allows avoiding issues that arise when applying quantum field theory to tachyons.

**References**

1. Recami, E., Classical Tachyons and Possible Applications **1985**, *Il Nuovo Cimento*, 9(6), 587-593.
2. Puscher, E.A.., Faster-Then-Light Particles: a Review of Tachyon Characteristics, The Rand Publication Series, **1980**, U.S.A.
3. Bilaniuk, O.M.P.; Surdashan, E.C.G.; Particles Beyond the Light Barrier **1969**, *Phys. Today*, 22(5), 43-51.
4. Craig, W.L.; Tachyons, Time Travel and Divine Omniscience **1988**, *J. of Phyl.*, 85, 135-150.
5. Recami, E., Casuality and Tachyons in Relativity, *Italian Studies in Philosophy of Science* **1980**, 47, 266.
6. Binaliuk, O.M.P.; Despande, V.K.; Surdashan, E.C.G.; Meta Relativity **1962**, *Am. J. Phys.*, 30(2), 718-723.
7. Feinberg, G.; Possibility of Faster Than Light Paricles **1967**, *Phys. Rev.*, 159(5), 1089-1105.
8. Tanaka, S., Theory of Matter with Superlight Velocity **1960**, *Prog. Theor. Phys.*, 24, 171-200.
9. Hamamoto, S., Quantization of Free Tachyon Fields and Possibility of Superluminal Urbanyon Model **1972**, *Prog. Theor. Phys.*, 48, 1037-1051.
10. Perepelitsa, V.F.; Looking for a Theory of Faster Than Light Particles **2015**, arxiv:1407.3245v4 [Physics.gen-ph].
11. Zweibach, B., A First Course in String Theory **2009**, Cambridge University Press, U.K.
12. Sahoo, S., Kumar, M., Creation of Tachyon Particles from Black Hole **2012**, *New Adv. Phys.*, 6 (1), 35-39.
13. Lemke, H.; Covariant Kinematics for the Production of Spacelike Particles **1980**, *Phys. Rev. D*, 22(6), 1342-1344.
14. Majorana, E.; Relativistic Theory of Particles with Arbitrary Intrinsic Angular Momentum **1932**, *Il Nuovo Cimento*, 9, 335-344. – English Translation by C.A. Orzalesi in Technical Report, 792, University of Meryland (1968).
15. Nanni, L.; Quantum Theory of Half-integer Spin Free Particles from the Perspective of the Majorana Equation **2016**, arxiv:1603.05965 [Physics.gen-ph].
16. Bisiacchi, G., Budini, P., Calucci, G., Majorana Equations for Composite Systems **1967**, *Phys. Rev.*, 172 (5), 1508-1515.
17. Casalbuoni, R., Majorana and the Infinite Component Wave Equations **2006**, arxiv:0610252 [hep.th].
18. Park, M.I., Park, Y.J., On the Fundation of the Relativistic Dynamics with the Tachyon **1996**, *Il Nuovo Cimento*, 111 (11), 1333-1368.